\documentstyle[pra,aps]{revtex}
\title{Strengthening the Bell Theorem: conditions
to falsify local realism in an experiment} 

\author{Marek \.Zukowski$^1$, Dagomir Kaszlikowski$^2$, Adam
  Baturo$^3$ and Jan-{\AA}ke Larsson$^4$}

\address{$^1$Instytut Fizyki Teoretycznej i Astrofizyki
  Uniwersytet Gda\'nski, PL-80-952 Gda\'nsk, Poland,\\
  $^2$Studium Doktoranckie z Fizyki,
  Uniwersytet Gda\'nski, PL-80-952 Gda\'nsk, Poland,\\
  $^3$Wydzia{\l } Matematyki i Fizyki,
  Uniwersytet Gda\'nski, PL-80-952 Gda\'nsk, Poland,\\
  $^4$Matematiska Institutionen, Link\"opings Universitet, SE-581 83
  Link\"oping, Sweden.}
\begin{document}
\twocolumn
\tighten
\maketitle
\begin{abstract}
  The two-particle correlation obtained from the quantum state used in
  the Bell inequality is sinusoidal, but the standard Bell inequality
  only uses two pairs of settings and not the whole sinusoidal
  curve. The highest to-date visibility of an explicit model
  reproducing sinusoidal fringes is $2/\pi$. We conjecture from a
  numerical approach presented in this paper that the highest possible
  visibility for a local hidden variable model reproducing the {\it
    sinusoidal} character of the quantum prediction for the
  two-particle Bell-type interference phenomena is
  $\frac{1}{\sqrt{2}}$.  In addition, the approach can be applied
  directly to experimental data.
\end{abstract}
 \pacs{PACS numbers: 3.65.Bz, 42.50.Dv}

It is a common wisdom in the quantum optical community that the
threshold visibility of the sinusoidal two-particle interference
pattern beyond which the Bell inequalities are violated is (for the
case of perfect detectors) $\frac{1}{\sqrt{2}}$ (see, e.\ g.\ 
\cite{CS}). Most of the experiments exceed that limit (with the usual
``fair sampling assumption'') \cite{ASPECT}. Some difficulties to
reach this threshold were observed in the very early experiments
\cite{CS}, as well as in some recent ones involving novel techniques.
Thus far, in atomic interferometry EPR experiments \cite{HAGLEY} and
for the phenomenon of entanglement swapping \cite{ENTSW}, the
resulting visibility is less than the magic $71\%$.

It is also well known that the Clauser-Horne inequality and the CHSH
inequality are not only necessary conditions for the existence of
local realistic models but also sufficient \cite{FINE} (in the case of
the CHSH inequality, this requires some simplifying assumptions
\cite{GARGMERMIN}). However, the sufficiency proofs used involve only
{\em two pairs of settings} of the local macroscopic parameters (e.g.
orientations of the polarizers) that define the measured local
observables. Thus, the constructions are valid for precisely those
settings and nothing more, and there is no guarantee that the models
can be extended to more settings.  Consequently one may ask what is
the maximal visibility for a model applicable to {\em all possible
  settings} of the measuring apparata, that returns {\em sinusoidal}
two particle interference fringes. It is already known that for
perfect detectors, this value cannot be higher than
$\frac{1}{\sqrt{2}}$ or lower than $2/\pi$ (this is the visibility of
the recent {\it ad hoc} model by Larsson \cite{LARSSON}; for earlier
models returning visibilities of $50\%$ see e.g. \cite{WODKIEWICZ}).

The knowledge of the maximal visibility of sinusoidal two-particle
fringes in a Bell-type experiment that still can be fully modeled in a
local realistic way, may help us to distinguish better between `local'
and `nonlocal' density matrices.  For two two-state systems one can
find precise conditions which have to be satisfied by density matrices
describing the general state, pure or mixed, of the full system, that
enable violation of the CHSH inequalities \cite{HOR}. States
fulfilling such conditions are often called ``nonlocal''. However,
since the CHSH inequality is necessary and sufficient only for two
pairs of settings, it is not excluded {\it a priori} that some states
that satisfy such inequalities for all possible sets of two pairs of
local dichotomic observables, nevertheless give predictions that in
their entirety cannot be modeled by local hidden variables. Such
models must first of all reproduce the full continuous sinusoidal
variation of two-particle interference fringes, as well as the other
predictions.

It is clear that the full solution of the question would require a
construction, or a proof of existence, of local hidden variable models
which return sinusoidal fringes of the maximal possible visibility,
that are applicable for all possible settings of the measuring
apparata.  Since this seems to be very difficult, we chose a numerical
method of pointwise approximation at a finite number of settings at
each side of the experiment. Due to exponential growth of the
computation time when the number of settings increases, we managed to
reach up to 9 settings on each side, i.e. up to 81 measurements points
(which due to a certain symmetry, about which we will say more later,
effectively can be transformed into $18\times18=324$ points).  The
exponential growth hinders any substantial increase in this
number\cite{comm1}. Such numerical models cannot give a definite
answer concerning the critical visibility of sinusoidal fringes,
however our calculations enable us to put forward a strong conjecture
that this value must be indeed ${1\over\sqrt2}$ (see below).

Experimentally our problem can be formulated in the following way: the
two particle state produced by the source does not allow for single
particle interference, and in the experiment less-than-perfect
two-particle fringes are obtained, due to some fundamental limitations
(like those present in the case of entanglement swapping, e.g.
\cite{ENTSW}) or due to imperfections of the devices. What is the
critical two particle interference visibility beyond which the
observed process falsifies local realism? We shall ask these questions
assuming, for simplicity, perfect detection efficiencies, which is
possible theoretically, and experimentally thus far amounts to the
usual ``fair sampling assumption''.  

Furthermore, the problem may be investigated without the use of the
assumption that the observed fringes are of a sinusoidal nature even
though two observed two-particle fringes in experiments with high
photon counts follow almost exactly the sinusoidal curves
\cite{ASPECT}. In experiments with lower count rates, still with
relatively good level of confidence the recorded data have
approximately the same character, and it is customary to fit them with
sinusoidal curves. It is now a standard procedure to perform the
two-particle interference experiments by recording many points of the
interference pattern, rather than stabilizing the devices at
measurement settings appropriate for the best violation of some Bell
inequality.  Further, in some of the experiments, e.g. those involving
optical fiber interferometers it is currently not possible to
stabilize the phase differences and what is observed is just the
interference pattern changing in time, and the visibility of the
sinusoidal two-particle fringes is used as the critical parameter
\cite{TAPSTER}.

Even though the numerical calculations presented here only reach
$18\times18$ points, this is more than enough in comparison to the
experimental data. The usual experimental scans rarely
involve more than 20 points. Further, our 
algorithm can be applied {\it directly to the
measurement data}, and in that way one can even avoid the standby
hypothesis that the fringes follow a sinusoidal pattern.  The
algorithm can directly answer the question: are the data compatible
with local realism or not?  Since physics is an experimental science,
the questions about Nature get their final answers solely in this way.

Let us now go to a formal treatment of the problem, and our numerical
solution of it. In a standard Bell-type experiment one has a source
emitting two particles, each of which propagates towards one of two
spatially separated measuring devices. The particles are described by
the maximally entangled state, e.g.,
\begin{eqnarray}
&|\Psi\rangle={1\over\sqrt2}(|+\rangle_{1}|-\rangle_{2}-
|-\rangle_{1}|+\rangle_{2}),&
\label{singlet}
\end{eqnarray}
where $|+\rangle_{1}$ is the state of the first particle with its spin
directed along the vector $\bbox{z}$ of a certain frame of reference
($-$ denotes the opposite direction), etc. We will now assume that the
measuring devices are Stern-Gerlach apparata, measuring the observable
$\bbox{n}\cdot\bbox{\sigma}$, where $n=a,b$ ($a$ for the first
observer, $b$ for the second one), $\bbox{n}$ is a unit vector
representing direction at which observer $n$ makes a measurement and
$\bbox{\sigma}$ is a vector the components of which are standard Pauli
matrices. The family of observables $\bbox{n}\cdot\bbox{\sigma}$
covers all possible dichotomic observables for a spin ${1\over 2}$
system, endowed with a spectrum consisting of $\pm 1$.

In each run of the experiment every observer obtains one of the two
possible results of measurement, $\pm 1$. The probability of obtaining
the result $m=\pm 1$ at the observer $a$, when measuring the
projection of the spin of the incoming particle at the direction
$\bbox{a}$, and the result $l=\pm 1$ at the observer $b$, when
measuring the projection of spin of the incoming particle at the
direction $\bbox{b}$ is equal to
\begin{eqnarray}
  &P_{QM}(m,l;\bbox{a},\bbox{b})={1\over4}(1-ml\bbox{a}\cdot\bbox{b}),&
  \label{predictions}
\end{eqnarray} 
while the probabilities of obtaining one of the results in the local
stations reveal no dependence on the local parameters,
$P_{QM}(m;\bbox{a})={1\over2}$ and $P_{QM}(l;\bbox{b})={1\over2}$.  In
a real experiment, however, one cannot expect that the observed
probabilities will follow (\ref{predictions}). Therefore, we will
allow that the interference pattern is of a reduced visibility.  In
such a case (\ref{predictions}) should be replaced by
\begin{eqnarray}
  &P_{QM}(m,l;\bbox{a},\bbox{b})={1\over4}(1-mlV\bbox{a}\cdot\bbox{b}),&
  \label{predictions2}
\end{eqnarray} 
where $0\leq V \leq 1$ stands for the visibility.

Now for the pointwise approximation. For $V=1$ the quantum prediction
for two-particle correlation function reads:
\begin{eqnarray}
  &E_{QM}(\bbox{a}, \bbox{b})
  = \sum_{m,l} mlP(m,l;{\bbox{a}},{\bbox{b}})=
  -\bbox{a}\cdot\bbox{b},&
\end{eqnarray}
and there is no single particle interference (i.e. the local results
are absolutely random). If one assumes that the unit vectors which
define the measured observables are always coplanar the correlation
function can be simplified to $E_{QM}(\alpha,
\beta)=-\cos{(\alpha-\beta)}$ (with the obvious definition of $\alpha$
and $\beta$).  Let us assume that in the experiment the observer at
the side $a$ chooses between, say, $N$ settings of the local apparatus,
denoted here by $\bbox{a}_i$ with $i=1,2,3, \ldots N$, and the other
observer, at side $b$, chooses between $M$ settings $\bbox{b}_j$, with
$j=1,2, \ldots, M$. The quantum correlation function for $V=1$ has at
these settings the following fixed numerical values:
$E_{QM}(\bbox{a}_i, \bbox{b}_j)=- \bbox{a}_i\cdot\bbox{b}_j.$ Thus, we
have a certain matrix of quantum predictions.  We can denote this
matrix by $\bf \hat{Q}$, with $Q_{ij}=E_{QM}(\bbox{a}_i, \bbox{b}_j)$.
For $V\leq 1$ the correlation function reduces to $VQ_{ij}$.

The Bell theorem is a statement on the impossibility of modeling
certain quantum predictions by local and realistic theories.  To make
the matters simpler let us use the local hidden variable (LHV)
formalism.  Within such a formalism the correlation function must have
the following structure
\begin{equation} 
  E_{LHV}(\bbox{a}_i, \bbox{b}_j)= 
  \int d\lambda\rho(\lambda)A(\bbox{a}_i,\lambda)B(\bbox{b}_j,\lambda), 
\end{equation}
where for dichotomic measurements $A(\bbox{a}_i,\lambda)=\pm1$ and
$B(\bbox{b}_j,\lambda)=\pm1$, and they represent the values of local
measurements predetermined by the LHV's, denoted by $\lambda$, for the
specified local settings. This expression is an average over a certain
LHV distribution $\rho(\lambda)$ of certain {\it factorizable} (rank
1) matrices, namely those with elements given by
$M(\lambda)_{ij}=A(\bbox{a}_i,\lambda)B(\bbox{b}_j,\lambda)$.  The
symbol $\lambda$ may hide very many parameters.  However, since the
only possible values of $A(\bbox{a}_i,\lambda)$ and
$B(\bbox{b}_j,\lambda)$ are $\pm1$ there are only $2^{N}$ {\it
  different} sequences of the values of $(A(\bbox{a}_1,\lambda),
A(\bbox{a}_2,\lambda),..., A(\bbox{a}_N,\lambda))$, and only $2^{M}$
different sequences of $
(B(\bbox{b}_1,\lambda),B(\bbox{b}_2,\lambda),...,B(\bbox{b}_M,\lambda))$
and consequently they form only $2^{N+M}$ matrices $\hat{M}(\lambda)$.
Therefore the structure of LHV models of $E_{LHV}(\bbox{a}_i,
\bbox{b}_j)$ reduces to discrete probabilistic models involving the
average of all the $2^{N+M}$ matrices $\hat{M}(\lambda)$.  In other
words, the LHV's can be replaced, without any loss of generality, by a
certain pair of variables $k$ and $l$ that have integer values
respectively from $1$ to $2^{N}$ and from $1$ to $2^{M}$.  To each $k$
we ascribe one possible sequence of the possible values of
$A(\bbox{a}_i,\lambda)$, denoted from now on by $A(\bbox{a}_i,k)$,
similarly we replace $B(\bbox{b}_j,\lambda)$ by $B(\bbox{b}_j,l)$.
With this notation the possible LHV models of the correlation function
$E_{LHV}(\bbox{a}_i, \bbox{b}_j)$ acquire the following simple form
\begin{equation} 
  E_{LHV}(\bbox{a}_i, \bbox{b}_j)=
  \sum_{k=1}^{2^N}\sum_{l=1}^{2^M}p_{kl}A(\bbox{a}_i,k)B(\bbox{b}_j,l),
  \label{MODEL}
\end{equation}
with, of course, the probabilities satisfying $p_{kl}\geq0$ and
$\sum p_{kl}=1$.

The special case that we study here enables us to simplify the
description further.  To satisfy the additional requirement that the
LHV model returns the quantum prediction of equal probability of the
results at the local observation stations, that $P(l;\bbox{a})=
P(m;\bbox{b})={1\over 2}$, one can use the following observation.
For each $k$, there must exist a $k'\neq k$ with the property that
$A(\bbox{a}_i,k')=-A(\bbox{a}_i,k)$, and similarly for each $l$, there
must exist an $l'\neq l$ for which
$B(\bbox{b}_j,l')=-B(\bbox{b}_j,l)$. Then
$A(\bbox{a}_i,k)B(\bbox{b}_j,l)=A(\bbox{a}_i,k')B(\bbox{b}_j,l')$, and
thus they give exactly the same matrix of LHV predictions. By assuming
$p_{kl}=p_{k'l'}$ the property of total randomness of local results
will always be reproduced by the LHV models, and the generality will
not be reduced since the contributions of $p_{kl}$ and $p_{k'l'}$ to
(\ref{MODEL}) cannot be distinguished. Of course in the actual
computer calculations of the correlation function we take only one
representative of the two pairs, reducing in this way the number of
probabilities and matrices of LHV predictions in (\ref{MODEL}) by a
factor of two.

Another reduction by a factor of four is given by the fact that in the
coplanar case, the choice of the settings may be limited on each side
to ranges not greater than $\pi$ (i.e. $\phi \leq \alpha_{i} \leq \phi
+\pi, \psi \leq\beta_{j} \leq \psi +\pi$). This is due to the simple
observation that a model of the type (\ref{MODEL}) once established
for such settings can be easily extended to settings
$\alpha_{i}^{'}=\alpha_{j}+\pi, \beta_{j}^{'}=\beta_{j}+\pi$ by
putting $A(\alpha_{i}^{'},l)=-A(\alpha_{i},l)$ and
$B(\beta_{j}^{'},l)=-B(\beta_{j},l)$ (nevertheless, some scans with
wider ranges were performed).

The conditions for LHV's to reproduce the quantum prediction with a
final visibility $V$ can be simplified to the problem of maximizing a
parameter $V$ for which exists a set of $2^{N+M}$ probabilities,
$p_{kl}$, such that 
\begin{equation}
  \sum_{k=1}^{2^N}\sum_{l=1}^{2^M}p_{kl}A(\bbox{a}_i,k)B(\bbox{b}_j,l)
  =VQ_{ij}.
  \label{MODEL1}
\end{equation}
This is a typical linear optimization problem in which we have more
unknowns than conditions and for which many good algorithms exist.
Therefore the simplest method to solve it is to use the standard
method of linear programming.  The core of the algorithm is a
procedure which finds the maximum of a linear function within the
given constraints.  In our case the constraints are the $N\times M$
equations (\ref{MODEL1}), and the condition that $\sum p_{kl}=1$. Our
unknowns are all $p_{kl}$ and $V$ (all nonnegative). We can treat them
as points in $2^{N+M}+1$ dimensional space. The constraints given by
(\ref{MODEL1}) define some subset of this space. On this subset we use
the trivial linear function $f(p_{kl},V)=V$ as our goal function (for
clarity, our function depends only on the variable $V$) and we search
for its maximum.

In order to apply this method to experimental results, replace
$Q_{ij}$ in (\ref{MODEL1}) by the measured values
$E^{\text{exp}}_{ij}$, and perform the same task. If the critical $V$
\cite{comm2} returned by the program is less that $1$, the data cannot
be reproduced by any LHV model.  Note that one even does not have to
know what the settings are(!).  This type of approach may be useful
especially when one is not able to stabilize the interferometers at
settings which are required for some Bell inequality, but there are
data at many other settings.

For the case of coplanar settings we have checked lots of
``interesting'' combinations of the settings of the apparata at each
side (e.g. $N\times N$ problems , $2\leq N \leq 9$, with evenly spaced
settings, etc. \cite{EXAMPLE}).  The main interesting {\it generic} feature of 
our
results is the following one.  Whenever among the coplanar settings
$\alpha_i$ and $\beta_j$ there is a subset $\alpha_{i_1}$,
$\alpha_{i_2}$ and $\beta_{j_1}$, $\beta_{j_2}$ such that for these
settings the CHSH inequality (equivalently CH inequality) is maximally
violated by the ideal quantum prediction, our optimum, the maximal
visibility reproduced by LHV's, is $V={1\over\sqrt2}$.  In all other
cases, namely for those settings without any such a subset, we have
obtained maximal visibilities describable by LHV's (usually, slightly)
{\it higher} than ${1\over\sqrt2}$. However with increasing number of
more or less evenly spaced settings this difference decreases. For
general measurement directions (i.e. those including non-coplanar
settings) all numerical scans follow the same pattern as for the
co-planar case. Additionally, we have checked 200000 randomly chosen
sets of $5\times5$ settings, and never a visibility lower than
${1\over\sqrt2}$ has been returned. All this strongly suggests that a
LHV model returning correlation function of the form
$\frac{1}{\sqrt{2}}\bbox{a}\cdot\bbox{b}$ in its entirety indeed
exists.

Furthermore, in the coplanar case, using carefully chosen equally
spaced settings more structure is introduced into the model, in the
form of two additional symmetries of the matrix $\hat{Q}$ (in addition
to the ones described above): ordinary matrix-transpose symmetry, and
constant diagonals. This reduces the rate of the exponential growth of
the calculation to a point where the problem is computable on a
standard PC in reasonable time, even for $13\times13$ (extendible to
$26\times26$) settings. Even in this case the returned visibilities
exhibit the behavior discussed above and always satisfy
$V\geq\frac{1}{\sqrt{2}}$.

Finally let us present application of our method to raw experimental
data. In a recent Bell-type experiment Weinfurter and Michler have
obtained the following matrix of results \cite{DATAWORK}
\begin{equation} 
  \widehat{E^{\text{exp}}}=\left[
    \begin{array}{ccc}
      -0.894 &  -0.061 &  0.761 \\
      -0.851 &  0.343 &  0.765 \\
      -0.625 &  0.688 &  0.516 \\
      -0.251 &  0.860 &  0.103 \\
      0.226 &  0.921 &  -0.389 \\
      0.530 &  0.651 &  -0.648 \\
      0.855 &  0.323 &  -0.832 \\
      0.852 &  -0.092 &  -0.843 \\
      0.785 &  -0.539 &  -0.638 \\
      0.397 &  -0.795 &  -0.253 \\
    \end{array}
  \right].
\end{equation} 
Our program gives the verdict that the values of all entries to the
matrix of results have to be reduced by the factor of $0.796$ to be
describable by local hidden variables.

In the recent long-distance EPR-Bell experiment the following set of
values of the correlation function was obtained \cite{WEIHS}:
\begin{equation} 
  \widehat{E^{\text{exp}}}=
  \left[
    \begin{array}{cc}
      0.960 &  -0.102 \\
      0.903 &  -0.375 \\
      0.733 &  -0.660 \\
      0.479 &  -0.809 \\
      0.191 &  -0.903 \\
      -0.120 &  -0.923 \\
      -0.429 &  -0.807 \\
      -0.666 &  -0.656 \\
      -0.842 &  -0.395 \\
      -0.951 &  -0.152 \\
      -0.953 &   0.171 \\
    \end{array}
  \right].
\end{equation}
This matrix has to be reduced by the factor of $0.7366$ to have a
local realistic description.

To conclude, the performed calculations enable us to put forward the
following conjecture, which is the main result of this letter:
sinusoidal two-particle fringes of visibility up to
$\frac{1}{\sqrt{2}}$ are describable by local realistic theories.  At
this stage we are not able to give an analytic proof of the above.
However, for finite sets of measurement points the results of data
analysis with the use of our program fully concur with this
hypothesis.  This implies, e.g. that one needs a re-run of the
entanglement swapping experiment, in order to show that this
phenomenon can lead to observable violations of local realism.

M\.Z was supported by the University of Gdansk Grant No
BW/5400-5-0264-9. DK was supported by the KBN Grant 2 P03B 096 15.
M\.Z thanks A. Zeilinger, H. Weinfurter and N. Gisin for 
discussions on the subject.


\begin{references}
  
\bibitem{CS} J.F. Clauser and A. Shimony (1978), Rep. Prog. Phys.,
  {\bf 41}, 1881.  
  
\bibitem{ASPECT} See, e.g., S.J. Freedman and J.S. Clauser, Phys. Rev.
  Lett. 28, 938 (1972); Aspect, A., P. Grangier and G. Roger, 1981,
  Phys. Rev. Lett. 47, 460; Z.Y. Ou and L. Mandel, Phys. Rev. Lett.
  {\bf61}, 50 (1988); J.G. Rarity and P.R. Tapster (1990), Phys. Rev.
  Lett., {\bf 64}, 2495; P. G. Kwiat, A. M. Steinberg and R. Y.
  Chiao, Phys. Rev. A {\bf 47}, R2472 (1993); T. B. Pittman, Y.H.
  Shih, A.V. Sergienko and M.H. Rubin, Phys. Rev. A {\bf 51}, 3495
  (1995); W. Tittel, J. Brendel, H. Zbinden, and N. Gisin, Phys. Rev.
  Lett. {\bf 81}, 3563 (1998), and ref. [15].

\bibitem{HAGLEY} E. Hagley, X. Maitre, G. Nogues, C. Wunderlich,
  M. Brune, J.M.\ Raimond and S. Haroche, Phys. Rev. Lett. {\bf 79}, 1
  (1997).  

\bibitem{ENTSW} M.\ \.Zukowski, A.\ Zeilinger, H.\ Weinfurter (1995),
  Ann.  N.Y.\ Acad.\ Sci., {\bf 755}, 91; J.-W.\ Pan, D. Bouwmeester,
  H.\ Weinfurter, and A.\ Zeilinger, Phys. Rev. Lett. {\bf 80}, 3891
  (1998).  

\bibitem{FINE} A.  Fine, J. Math. Phys. {\bf 23}, 1306 (1982).  

\bibitem{GARGMERMIN} A.  Garg and N.D. Mermin, Phys. Rev. D {35}, 3638
  (1987).  
  
\bibitem{LARSSON}J.-\AA. Larsson, Phys. Lett. A {\bf 256}, 245 (1999).  

\bibitem{WODKIEWICZ} C. Su and K. W\'odkiewicz, Phys. Rev. A {\bf 44},
 6097 (1991).  

\bibitem{HOR} R. Horodecki, P. Horodecki and M.  Horodecki,
  Phys. Lett. A {\bf 200}, 340 (1995).  
  
\bibitem{comm1} Recently A. Peres, quant-ph-9807017 has discussed a
  computer algorithm which searches for so-called Farkas vectors,
  which in turn define coefficients in generalized Bell-inequalities,
  the set of which is a sufficient and necessary condition for
  classical probabilistic model (here, essentially, local realistic)
  to reproduce a certain set of probabilities for pairs of
  experiments.  However, his method explodes numerically much much
  faster then ours.  Simply, our method is applied directly to a
  certain finite set of specified quantum prediction or experimental
  data.  Whereas inequalities based on the Farkas lemma apply to {\it
    all possible sets of data}.

\bibitem{TAPSTER} P. R.  Tapster, J. G. Rarity, and P. C.  M.  Owens,
  Phys. Rev. Lett., {\bf 73}, 1923 (1994)
  
\bibitem{comm2} In this case $V$ does not have the direct
  interpretation of visibility. Its value tells us by what factor the
  observed values of the correlation function have to be reduced, so
  that a local hidden variables model exists.

\bibitem{EXAMPLE}For example, for $\alpha=0^o, 10^o, 20^o,...,80^o$,
and $\beta=-45^o,-35^o,-25^o,...,+35^o$  the critical $V $ is $0.722536$. 
\bibitem{DATAWORK} H. Weinfurter, M.  Michler, private communication.
  The numbers give the values of the correlation function - there were
  three different setting on side A of the experiment and 27 settings
  on side B, however only 10 of them are shown here.  In the actual
  experiment only data from a pair of detectors were collected.  To
  obtain the matrix we used the usual assumption that
  $E(\alpha,\beta)=4P(+,+;\alpha, \beta)-1.$ We have also renormalized
  the numbers of photon pairs counted, so that the average of the
  counts over approximately two periods of settings at side B
  represents the probability of $\frac{1}{4}$ (in concurrence with the
  quantum prediction). To this end we used data for all 27 settings on
  side B.
  
\bibitem{WEIHS} G. Weihs, T. Jennewein, Ch.  Simon, H. Weinfurter and
  A. Zeilinger, Phys. Rev. Lett. {\bf 81}, 5039 (1998).  The entries
  to the matrix were provided by G. Weihs (private communication).

\end{references}
\end{document}